\documentclass[12pt]{amsart}
\usepackage[english]{babel}
\usepackage{amsmath,amsthm,amsfonts,amssymb,epsfig,color,ulem,hyperref}
\usepackage[left=1in,top=1in,right=1in]{geometry}

\newtheorem{thm}{Theorem}[section]

\newtheorem{df}[thm]{Definition}

\newcommand{\Z}{\mathbb{Z}}

\begin{document}

\title{Ground state stability in two spin glass models}
%
%

\author[L.-P.~Arguin]{L.-P.~Arguin}
 \address{L.-P.~Arguin\\
 Department of Mathematics\\
 City University of New York, Baruch College and Graduate Center\\
 New York, NY 10010}
\email{louis-pierre.arguin@baruch.cuny.edu}

\author[C.M.~Newman]{C.M.~Newman}            
 \address{C.M.~Newman\\ 
 Courant Institute of Mathematical Sciences\\
 New York, NY 10012 USA\\
  and NYU-ECNU Institute of Mathematical Sciences at NYU Shanghai\\
  3663 Zhongshan Road North, Shanghai 200062, China}
\email{newman@cims.nyu.edu}

\author[D.L.~Stein]{D.L.~Stein}            
 \address{D.L.~Stein\\ 
 Department\ of Physics and Courant Institute of Mathematical Sciences\\
  New York University\\
 	 New York, NY 10003, USA\\
	 and NYU-ECNU Institutes of Physics and Mathematical Sciences at NYU Shanghai\\
	  3663 Zhongshan Road North\\
	  Shanghai, 200062, China\\
	  and Santa Fe Institute, 1399 Hyde Park Rd., Santa Fe, NM 87501 USA}
\email{daniel.stein@nyu.edu}

\begin{abstract}
An important but little-studied property of spin glasses is the stability of their ground states to changes in one or a finite number of couplings.
It was shown in earlier work that, if multiple ground states are assumed to exist, then fluctuations in their energy differences --- and therefore the possibility of multiple ground states ---
are closely related to the stability of their ground states. Here we examine the stability of ground states in two models, one of which is presumed to have a ground state structure that is
qualitatively similar to other realistic short-range spin glasses in finite dimensions.
\end{abstract}

\maketitle

\section{Introduction and definitions}
\label{sec:intro}

Vladas was a remarkable mathematician, collaborator, colleague and friend: often exciting, always interesting, sometimes frustrating but never boring.
We will miss him greatly, but are confident that his memory will survive for a very long time.

Although he never worked directly on spin glasses himself, Vladas maintained a longstanding interest in the problem, and we enjoyed numerous discussions
with him about possible ways of proving nonuniqueness of Gibbs states, energy fluctuation bounds, overlap properties, and many other open problems.
In this paper we discuss another aspect of spin glasses, namely ground state stability and its consequences, a topic we think Vladas would have enjoyed.

The stability of a spin glass ground state can be defined in different ways; here we will adopt the notion introduced in~\cite{NS2D00,NS2D01} and further developed and exploited
in~\cite{ADNS10,ANS19}. For specificity consider the Edwards-Anderson (EA) Ising model~\cite{EA75} in a finite volume $\Lambda_L=[-L,L]^d\cap\Z^d$ centered at the origin, with Hamiltonian
\begin{equation}
\label{eq:EA}
H_{\Lambda, J}(\sigma)=-\sum_{\langle xy\rangle\in E(\Lambda)}J_{xy} \sigma_x\sigma_y, \qquad \sigma\in \{-1,1\}^{\Lambda} \ ,
\end{equation}
where $E(\Lambda)$ denotes the set of nearest-neighbor edges $\langle xy\rangle$ with both endpoints in $\Lambda$.  The couplings~$J_{xy}$ are i.i.d.~random variables sampled from a continuous distribution $\nu(dJ_{xy})$, which for specificity we take to be~$\mathcal{N}(0,1)$.  If periodic or free boundary conditions are imposed, ground states appear as spin-reversed pairs.

For any fixed $\Lambda_L$ and accompanying boundary condition, the ground state configuration (or the ground state pair if the boundary condition has spin-flip symmetry) is denoted by $\alpha$. One may now ask the question, how does the lowest-energy spin configuration~$\alpha$ change when one selects an arbitrary edge $b_0$ and varies its associated coupling $J_0$  from $-\infty$ to $+\infty$?  If $J_0$ is satisfied, increasing its magnitude will only increase the stability of $\alpha$ and so the lowest-energy spin configuration pair is unchanged. However, if its magnitude is decreased, $\alpha$ becomes less stable, and there exists a specific value $J_c$ for which a cluster of connected spins (which we shall refer to as the ``critical droplet'') will flip, leading to a new ground state pair~$\alpha'$.  The same result follows if $J_0$ is unsatisfied and its magnitude is then increased.

More precisely, note that a ground state pair (hereafter GSP) is a spin configuration such that the energy~$E_{\partial\mathcal{D}}$  of any closed surface $\partial\mathcal{D}$  in the dual lattice satisfies the condition
\begin{equation}
\label{eq:gspstable}
E_{\partial\mathcal{D}}=\sum_{\langle xy\rangle\in\partial\mathcal{D}} J_{xy}\sigma_x\sigma_y>0\, .
\end{equation}
The critical value $J_c$ corresponds to the coupling value at which $\sum_{\langle xy\rangle\in\partial\mathcal{D}} J_{xy}\sigma_x\sigma_y=0$ in $\alpha$ for a {\it single\/} closed surface whose boundary passes through $b_0$, while all other such closed surfaces satisfy~(\ref{eq:gspstable}). The cluster of spins enclosed by the zero-energy surface $\partial\mathcal{D}_c(b_0,\alpha)$ is denoted the ``critical droplet'' of $b_0$ in the GSP~$\alpha$. Because the couplings are~i.i.d., $J_c$ depends on~$\alpha$ and all coupling values {\it except\/} that associated with $b_0$; that is, the critical value $J_c$ is independent of $J_0$. For a fixed coupling realization in which $J(b_0)=J_0$, we can therefore define the {\it flexibility\/}~$\mathcal{F}_{b_0,\alpha}$ of $b_0$ in $\alpha$ as 

\begin{equation}
\label{eq:flex}
\mathcal{F}_{b_0,\alpha}=|E_{\partial\mathcal{D}_c(b_0,\alpha)}(J_c)-E_{\partial\mathcal{D}_c(b_0,\alpha)}(J_0)|\, .
\end{equation}
Because the couplings are i.i.d.~and drawn from a continuous distribution, all flexibilities are strictly positive with probability one.

The presentation just given is informal; a complete discussion requires use of the {\it excitation metastate\/}~\cite{NSrc01,ADNS10,AD11,ANSW14} which we omit here for the sake of brevity. A precise definition of the above concepts and quantities can be found in~\cite{ANS19}.

The concepts of critical droplets and flexibilities for a particular GSP in a fixed coupling realization provide a foundation for quantifying (at least one version of) the stability of a given ground state. From an energetic standpoint, one can consider, e.g., the distribution of flexibilities over all bonds. One can also approach the problem from a geometric perspective, by considering the sizes and geometries of the critical droplets associated with each of the bonds. This latter approach has recently proved to be useful, in that the distribution of critical droplet sizes has been shown~\cite{ANS19} to be closely related to the energy fluctuations associated with collections of incongruent GSP's, i.e., GSP's whose mutual interfaces comprise a positive fraction of all edges in the infinite-volume limit~\cite{HF87a,FH87b}.

The problem with this approach, for now at least, is that there currently exist no tools or insights into determining ground state stability properties in ordinary EA~models. In this paper we discuss two models, one of which should belong in the same universality class as the ordinary EA model, in which some information on these properties {\it can\/} be determined.

\section{The highly disordered model}
\label{sec:hd}

\subsection{Definition and properties}
\label{subsec:def}

The highly disordered model was introduced in~\cite{NS94,NS96} (see also~\cite{BCM94}). It is an EA-type model defined on the lattice $\mathbb{Z}^d$ whose Hamiltonian in any $\Lambda\subset\mathbb{Z}^d$  is still given by~(\ref{eq:EA}); the difference is that now the coupling distribution is volume-dependent even though the coupling values remain i.i.d.~for each~$\Lambda$. The idea is to ``stretch out'' the coupling distribution so that, with probability one, in sufficiently large volumes each coupling magnitude occurs on its own scale. More precisely, each coupling magnitude is at least twice as large as the next smaller one and no more than half as large as the next larger one.

While there are many possibilities for the volume-dependent distribution of couplings, we have found it convenient to work with the following choice.  First, we associate two new i.i.d. random variables with each edge $\langle xy\rangle$: $\epsilon_{xy}=\pm 1$ with equal probability and $K_{xy}$ which is uniformly distributed in the closed interval~$[0,1]$. We then define the set of couplings $J_{xy}^{(L)}$ within $\Lambda_L$ as follows:
\begin{equation}
\label{eq:hdcouplings}
J_{xy}^{(L)}=c_L\epsilon_{xy}e^{-\lambda^{(L)}K_{xy}}\, ,
\end{equation}
where $c_L$ is a scaling factor chosen to ensure a sensible thermodynamic limit (but which plays no role in ground state selection), and $\lambda^{(L)}$ is a scaling parameter that grows quickly enough with~$L$ to ensure that the condition described at the end of the previous paragraph holds. It was shown in~\cite{NS96} that~$\lambda^{(L)}\ge L^{2d+1+\delta}$ for any $\delta>0$ is a sufficient condition.

We should emphasize that although the couplings $J_{xy}^{(L)}$ depend on $L$, the $K_{xy}$'s and $\epsilon_{xy}$'s do not; hence there is a well-defined infinite-volume notion of ground states for the highly disordered model on all of  $\mathbb{Z}^d$. This
is the subject of the theorem in the next subsection.

When the highly disordered condition is satisfied, the problem of finding ground states becomes tractable; in fact, a simple greedy algorithm provides a fast and efficient way to find the exact ground state in a fixed volume with given boundary conditions~\cite{NS94,NS96}. Moreover, the ground state problem can be mapped onto invasion percolation~\cite{NS94,NS96} which facilitates analytic study. It was further shown in~\cite{NS94,NS96} that in the limit of infinite volume the highly disordered model has a single pair of ground states in low dimension, and uncountably many pairs in high dimension. The crossover dimension was found to be six in~\cite{JR10}. It should be noted that this result, related to the minimal spanning tree, is rigorous only in dimension two (or in quasi-planar lattices~\cite{NTW17}).

The details of ground state structure in the highly disordered model have been described at length in~\cite{NS94,NS96} (see also~\cite{NS8D01,JR10}) and are not recounted here. In this contribution we present a new result, concerning the ground state stability of the highly disordered model, where it turns out that this model is tractable as well. The result we prove below is twofold: first, that with probability one {\it all\/} couplings have finite critical droplets in any ground state, and moreover this result is dimension-independent, and therefore independent of ground state pair multiplicity. We caution, however, that (as with all other results pertaining to this model) these results may be confined to the highly disordered model alone and have not been shown to carry over to the Edwards-Anderson or other realistic spin glass models. We will address this question more in the following section.

Before proceeding, we need to introduce some relevant properties and nomenclature pertaining to the highly disordered model. One of its distinguishing features --- and the central one for our purposes --- is the separation of all bonds into two distinct classes~\cite{NS94,NS96}. The first class, which we denote as {\it S1~bonds\/} are those that are satisfied in any ground state {\it regardless\/} of the sign of the coupling, i.e., that of~$\epsilon_{xy}$. These are bonds that are always satisfied, in every ground state. The remaining bonds, which we call~{\it S2\/}, are those in which a change of sign of their $\epsilon_{xy}$ value changes their status in any ground state from satisfied to unsatisfied or vice-versa. (Obviously, any unsatisfied bond in any ground state is automatically S2, but a satisfied bond could {\it a priori\/} be of either type.)

To make this distinction formal, we introduce the concept of {\it rank\/}:  In a given~$\Lambda_L$, the coupling with largest magnitude (regardless of sign) has rank one (this is the coupling with highest rank and the smallest value of $K_{xy}$); the coupling with the next largest magnitude has rank two; and so on. We then define an S1 bond as follows:

{\df A bond $\langle xy\rangle$ is S1 in~$\Lambda_L$ if it has greater rank than at least one coupling in any path (excluding the bond itself) that connects its two endpoints ${\bf x}$ and ${\bf y}$.}

\bigskip

In the above definition, we need to specify what is meant by a path if each endpoint connects to a point on the boundary. For fixed boundary conditions of the spins on $\partial\Lambda_L$, all points on the boundary are considered connected (often called wired boundary conditions), so disjoint paths from ${\bf x}$ to~$\partial\Lambda_L$ and ${\bf y}$ to $\partial\Lambda_L$ are considered as connecting ${\bf x}$ and ${\bf y}$. It follows from the definition that for wired boundary conditions an S1 bond in $\Lambda_L$ remains S1 in all larger volumes. These bonds completely determine the ground state configurations, while the S2 bonds play no role.\footnote{One can define S1 and S2 bonds for the EA model as well, in the sense that the EA model also possesses bonds that are satisfied in every ground state (though the precise definition used above no longer applies). There are of course far fewer of these in the EA model than in the highly disordered model, and there is no evidence that these ``always satisfied'' bonds play any special role in ground state selection in that model. (One possibly relevant result, that unsatisfied edges don't percolate in the ground state, was proved in~\cite{Berger17}.) } For free boundary conditions, a path connects ${\bf x}$ and ${\bf y}$ only if it stays entirely within $\Lambda_L$, never touching the boundary; i.e., points on the boundary are no longer considered connected. For periodic or antiperiodic boundary conditions, boundary points are considered connected to their image points but to no others. The reasons for these distinctions are provided in~\cite{NS96}, but are not relevant to the present discussion and are presented only for completeness.

It was proved in~\cite{NS94} and~\cite{NS96} that the set of all S1 bonds forms a union of trees, that every site belongs to some S1 tree, and that every S1 tree touches the boundary of $\Lambda_L$. The S1 bonds in a given $\Lambda_L$ in some fixed dimension form either a single tree or else a union of disjoint trees. Although not immediately obvious, it was proved in~\cite{NS94,NS96} that the tree structure has a natural infinite volume limit, and moreover every tree is infinite.  Moreover, a result from Alexander~\cite{Alexander95}, adapted to the current context, states that if the corresponding independent percolation model has no infinite cluster
at $p_c$, then from every site there is a single path to infinity along S1 edges; i.e., there are no doubly-infinite paths.  It is widely believed that in independent percolation there is no infinite cluster at $p_c$ in any dimension, but this has not yet been proven rigorously for $3\le d\le 10$.

Finally, combined with results of Jackson and Read~\cite{JR10}, we have that below six dimensions there is a single S1 tree spanning the sites of~$\mathbb{Z}^d$ (corresponding to a single pair of ground states), while above six dimensions that are infinitely many trees (corresponding to an uncountable infinity of ground states).

\subsection{Ground state stability in the highly disordered model}
\label{subsec:stability}

Unlike in realistic spin glass models, the ground state structure in the highly disordered model can be analyzed and understood in great detail. This allows us to solve other, related properties of the model, in particular some of the critical droplet properties that have so far been inaccessible in most other spin glass models. In particular we can prove the following result:

{\thm In the highly disordered model on the infinite lattice~$\mathbb{Z}^d$ in any $d$, if there is no percolation at $p_c$ in the corresponding independent bond percolation model, then for a.e.~realization of the couplings, any ground state~$\alpha$, and any bond~$b_0$, the critical droplet boundary~$\partial\mathcal{D}_c(b_0,\alpha)$ is finite.  Correspondingly, in finite volumes $\Lambda_L$ with sufficiently large $L$, the size of the droplet is independent of $L$.}

\medskip

\noindent {\it Remark.} As noted above, it has been proved that there is no percolation at $p_c$ in the corresponding independent bond percolation model in all dimensions except $3\le d\le 10$, but it is widely believed to be true in all finite dimensions. Theorem~2.2 does not specify the distribution of critical droplet boundary sizes, which is potentially relevant especially for larger critical droplets, although integrability of the distribution requires a weak upper bound falloff such as~$O(L^{-(1+\epsilon)})$, $\epsilon>0$, for large~$L$.

\medskip

\noindent {\it Proof.\/} Choose an arbitrary S1 bond and a volume sufficiently large so that the tree it belongs to has the following property: The branch emanating from one of its endpoints (call it $x_1$) touches the boundary (on which we apply fixed boundary conditions) and the branch emanating from the other endpoint ($x_2$) does not. This remains the case as the boundary moves out to infinity: for any S1 bond and a sufficiently large volume, this is guaranteed to be the case by the result of Alexander mentioned above~\cite{Alexander95}.

We use the fact, noted in Sect.~\ref{sec:intro}, that as the coupling value of any bond varies from $-\infty$ to $+\infty$ while all other couplings are held fixed, there is a single, well-defined critical point at which a unique cluster of spins, i.e., the critical droplet, flips, changing the ground state. (This is true regardless of whether one is considering a finite volume with specified boundary condition or the infinite system.) Now keep the magnitude of the S1 bond fixed but change its sign. Because the S1 bond must still be satisfied, this must cause a droplet flip, which as noted above must be the critical droplet.

Now consider the state of the spins at either endpoint of the bond. Suppose that originally the bond was ferromagnetic, and the spins at $x_1$ and $x_2$ were both $+1$. After changing the sign of the coupling, the spin at $x_1$, remains $+1$ (because it is connected to the boundary, as explained in the first  paragraph of this proof) while the spin at $x_2$ is now $-1$. This must simultaneously flip all the spins on the branch of the tree connected to $x_2$. This is a finite droplet and as the chosen $S1$ bond was arbitrary, the critical droplet of any $S1$ bond likewise must be finite.

Consider now an S2 bond. Without changing its sign, make its coupling magnitude sufficiently large (or equivalently, its $K_{xy}$~value sufficiently small) so that it becomes S1. (This will cause a rearrangement of one or more trees, but it can be seen that any corresponding droplet flip must also be finite.) Now change the sign of the coupling. The same argument as before shows that the corresponding droplet flip is again finite. But given that the critical droplet corresponding to a given bond is unique, this was also the critical droplet of the original S2 bond. $\diamond$

\section{The strongly disordered model}
\label{sec:sd}

Although the highly disordered model is useful because of its tractability, it is clearly an unrealistic model for laboratory spin glasses. This leads us to propose a related model that, while retaining some of the simplifying features of the highly disordered model, can shed light on the ground state properties of realistic spin glass models. We will refer to this new model as the {\it strongly disordered model\/} of spin glasses.

The main difference between the two models is that in the strongly disordered model the couplings have the same distribution for all volumes. This is implemented by removing the volume dependence of the parameter~$\lambda$:

{\df The strongly disordered model is identical to the highly disordered model but with Eq.~(\ref{eq:hdcouplings}) replaced by
\begin{equation}
\label{eq:sdcouplings}
J_{xy}=\epsilon_{xy}e^{-\lambda K_{xy}}
\end{equation}
with the constant $\lambda\gg 1$ independent of $L$.}

\bigskip

In the strongly disordered model, the condition that every coupling value is no more than half the next larger one and no less than twice the next smaller one breaks down in sufficiently large volumes. This can be quantified: let $g(\lambda)={\rm Prob}(1/2\le e^{-\lambda K_{xy}}/e^{-\lambda K_{x'y'}}\le 2)$. That is, $g(\lambda)$ is the probability that any two arbitrarily chosen bonds have coupling values that do {\it not\/} satisfy the highly disordered condition. A straightforward calculation gives $g(\lambda)=2\ln 2/\lambda$. 

The strongly disordered model carries two advantages. On the one hand, its critical droplet properties are analytically somewhat tractable given its similarity to the highly disordered model. On the other hand, since its coupling distribution is i.i.d.~with mean zero and finite variance, and not varying with~$L$, we expect global properties such as ground state multiplicity to be the same as in other versions of the EA spin glass with more conventional coupling distributions.

{\thm If there is no percolation at $p_c$ in the corresponding independent bond percolation model, then in the strongly disordered model, the critical droplet of an arbitrary but fixed bond is finite with probability approaching one as $\lambda\to\infty$.}

\bigskip

{\bf Proof.}  Consider a fixed, infinite-volume ground state on $\mathbb{Z}^d$; this induces a (coupling-dependent and ground-state-specific) spin configuration on the boundary $\partial{\Lambda_L}$ of any finite volume $\Lambda_L\subset\mathbb{Z}^d$.

Consider an arbitrary edge $\{x_0,y_0\}$. Let $R$ denote the (random) smallest value in the invasion/minimal spanning forest model on $\mathbb{Z}^d$, defined by the i.i.d.~$K_{xy}$ (but with $K_{x_0y_0}$ set to zero, for convenience of the argument) such that one of the branches from $x_0$ or $y_0$ is contained within a cube of side length~$2R$ centered at $\{x_0,y_0\}$.  By the result of Alexander~\cite{Alexander95}  mentioned earlier, $R$ is a finite random variable (depending on the $K_{xy}$'s) if there is no percolation at $p_c$ in the corresponding independent bond percolation model.

Now choose a deterministic $\Lambda_L$ and consider the two events:  (a) $A_L=\{R<L/2\}$ and (b) $B_L=\{$the highly disordered condition is valid in the cube of side~$L$ centered at $\{x_0,y_0\}\}$. Because $R$ is a finite random variable, ${\rm Prob}(A_L)$ can be made arbitrarily close to one for $L$~large. Moreover, from the definition of the highly disordered model ${\rm Prob}(B_L)$ can also be made close to~one by choosing $\lambda$ large (for the given $L$). Specifically, let $P_0$ denote the probability that the critical droplet of $\{x_0,y_0\}$ is finite. Then $P_0\ge 1-\epsilon$ if  ${\rm Prob}(R>L/2)+CL^{2d}/\lambda\le\epsilon$ for some fixed $C>0$. But for any $\epsilon>0$ one can choose a sufficiently large $L$ so that ${\rm Prob}(R>L/2)\le\epsilon/2$, and then choose $\lambda$ such that $CL^{2d}/\lambda\le\epsilon/2$. The result then follows. $\diamond$

Theorem~3.2 sets a strong upper bound~$O(\lambda^{-1})$ on the fraction of bonds that might~{\it not\/} have a finite critical droplet. We do not yet know whether this gap can be closed in the sense that the strongly disordered model might share the property that {\it all\/} bonds have finite critical droplets. It could be that this is not the case, but that if the number of bonds with infinite critical droplets is sufficiently small, theorems analogous to those in~\cite{ANS19} can be applied. Work on these questions is currently in progress.

{\bf Acknowledgments.}  The research of L.-P.~A.~was supported in part by NSF CAREER DMS-1653602. The research of CMN was supported in part by NSF Grant DMS-1507019.


\end{document}